\documentclass[prl,twocolumn,showpacs,superscriptaddress,preprintnumbers,amsmath,amssymb,floatfix]{revtex4}

\usepackage{graphicx}
\usepackage{color}

\begin{document}

\title{Multi-plasmon absorption in graphene}

\author{Marinko Jablan}
\email{mjablan@phy.hr}
\affiliation{ICFO-Institut de Ciencies Fotoniques, Mediterranean Technology Park, 08860 Castelldefels (Barcelona), Spain}
\affiliation{Department of Physics, University of Zagreb, 10000 Zagreb, Croatia}

\author{Darrick E. Chang}
\affiliation{ICFO-Institut de Ciencies Fotoniques, Mediterranean Technology Park, 08860 Castelldefels (Barcelona), Spain}

\date{\today}

\begin{abstract}
We show that graphene possesses a strong nonlinear optical response in the form of multi-plasmon absorption, with exciting implications in classical and quantum nonlinear optics. Specifically, we predict that graphene nano-ribbons can be used as saturable absorbers with low saturation intensity in the far-infrared and terahertz spectrum. Moreover, we predict that two-plasmon absorption and extreme localization of plasmon fields in graphene nano-disks can lead to a plasmon blockade effect, in which a single quantized plasmon strongly suppresses the possibility of exciting a second plasmon. 
\end{abstract}

\maketitle

The field of nonlinear optics ranges from fundamental questions concerning light-matter interactions to exciting technological applications \cite{Boyd}.
However, usually very large field intensities are required to observe nonlinear effects. One is thus always looking for systems that will exhibit nonlinear phenomena at lower powers, with the ultimate limit being strong interactions between just two quanta of light~\cite{Chang2014}.
One possibility to increase nonlinear effects is to use the strong localization and enhancement of electromagnetic fields in the form of surface plasmon excitations~\cite{Kauranen2012}. 
In that regards, we note that graphene \cite{Novoselov2004} has been demonstrated to support extremely localized plasmons~\cite{Wunsch2006,Hwang2007,Jablan2009,Jablan2013,Ju2011,Yan2012,Chen2012,Fei2012}.
While optical nonlinearities in graphene have been studied by several authors \cite{Mikhailov2008,Wright2009,Ishikawa2010,Hendry2010,Sun2010,Yang2011, Mikhailov2011,Yao2012,Gullans2013,Manzoni2014,Cox2014,AlNaib2014}, here we predict a novel nonlinear effect in the form of multi-plasmon absorption. 
We also show how this effect leads to saturable absorption in graphene nano-ribbons at low input powers in the far-infrared and terahertz spectrum. Moreover, we predict that the extreme localization of plasmon fields in graphene nano-disks leads to such a strong two-plasmon absorption that it becomes nearly impossible to excite a second plasmon quanta in the system. This plasmon blockade effect would cause the nano-disk to behave essentially like a quantum two-level system, which is observable in its resonance fluorescence spectrum.

Graphene is a two-dimensional hexagonal lattice of carbon atoms \cite{Novoselov2004}. The low-energy band structure of graphene is described by Dirac cones with the electron dispersion $E_{n \bf k}=n\hbar v_F |{\bf k}|$, where $v_F=10^6$ m/s, and $n=\pm 1$ stands for the conduction (valence) band \cite{Wallace1954}. In its intrinsic form graphene is a zero-gap semiconductor; however, it can also be easily doped with free carriers and as such it supports plasmon modes \cite{Wunsch2006,Hwang2007,Jablan2009}. At low frequencies, one can get a rather accurate description of these modes by using a simple Drude conductivity,
\begin{equation}
\sigma(\omega)=\frac{e^2 E_F}{\pi \hbar^2} \frac{i}{\omega+i\gamma},
\label{Drude}
\end{equation}
where $E_F$ is the Fermi energy of graphene and $\gamma$ is a phenomenological damping rate that takes into account various decay channels like impurity or phonon scattering~\cite{Jablan2009}. The resulting plasmon dispersion is given by
\begin{equation}
q=\frac{2\pi\varepsilon_0\bar\varepsilon_r\hbar^2}{e^2 E_F}\omega^2,
\label{plasmon_dispersion}
\end{equation}
and we assume the average dielectric permittivity 
$\bar\varepsilon_r\approx 2.5$, which roughly corresponds to the case of graphene on SiO$_2$ substrate and air on top. This simple Drude model breaks down at large frequencies when plasmons become strongly damped by electron-hole excitations, which can be described by the Random Phase Approximation (RPA) \cite{Wunsch2006,Hwang2007,Jablan2009}. However, at low energies the Pauli principle blocks this decay channel and the plasmon is a long-lived excitation \cite{Wunsch2006,Hwang2007,Jablan2009,Jablan2013,Ju2011,Yan2012,Chen2012,Fei2012}. The resulting plasmon wavelength $\lambda_p=2\pi/q_p$ is about 100 times smaller than the free space wavelength $\lambda=2\pi c/\omega$, leading to the extreme localization of electromagnetic fields \cite{Jablan2009}.

An intuitive explanation of the strong nonlinearities associated with plasmons emerges by considering the typical doping levels in graphene. For an electron density of $n=10^{12}$ cm$^{-2}$, the distance between two electrons is $r_e=1/\sqrt{n\pi}=5.6$ nm, and so to observe some kind of nonlinear phenomena we need to compete with an intrinsic electric field of the order
\begin{equation}
E_e=\frac{e}{4\pi\bar\varepsilon_r\varepsilon_0 r_e^2}
\approx 2\cdot 10^7 \; \mbox{V/m}.
\label{E_field_intrinsic}
\end{equation}
This is significantly smaller than the characteristic field amplitude associated with nonlinear effects in atoms~\cite{Boyd}, given by the field between an electron and proton at a distance of a Bohr radius $a_B=0.5 \;\mbox{\AA}$. The electric field in that case is $E_{at}=5\cdot 10^{11}$ V/m, about 4 orders of magnitude larger than the field $E_e$ in graphene!

To see what happens to plasmons at such a field strength $E_e$, let us imagine a general case of plasma oscillations at frequency $\omega$ and wavevector $q$, which is accompanied by an electric field $E({\bf r},t)=\frac{1}{2}E_p e^{i{\bf q}\cdot{\bf r}-i\omega t}+c.c.$ in the plane of graphene.
When the plasmon field is small, it will induce a surface charge density $\rho_p=\frac{q\sigma(\omega)}{\omega}E_p$ that in turn creates an electric field $E_p=\frac{-i\rho_p}{2\bar\varepsilon_r\varepsilon_0}$, thus driving self-sustained charge density oscillations. However, at the field strength $E_p=E_e$, the induced charge density $|\rho_p|=e n/2$ will be comparable to the initial charge density $en$, and this simple picture of plasmons breaks down. We will in fact see that at this field $E_e$ there is a strong plasmon damping via the multi-plasmon absorption.

\begin{figure}
\centerline{
\mbox{\includegraphics[width=0.5\textwidth]{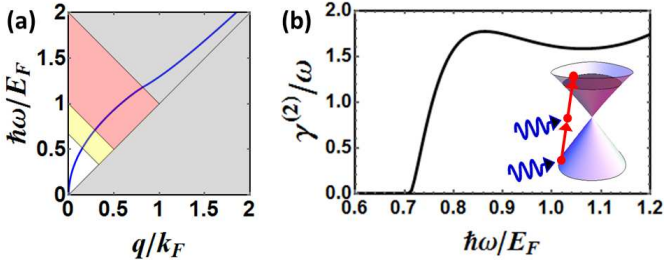}}
}
\caption{
(a) Plasmon dispersion relation. Gray area denotes the regime where a single plasmon can excite an e-h pair. Red area denotes the regime where this process is forbidden but plasmons can decay via two-plasmon absorption. Similarly, the yellow area denotes the regime of three-plasmon absorption.
(b) Two-plasmon damping rate $\gamma^{(2)}/\omega=F^{(2)}(\omega)$ for $E_p=E_e$, indicating that plasmons cannot oscillate when the plasmon field is equal to the intrinsic electric field. Inset shows the band structure of graphene and a two-plasmon absorption process.
}
\label{fig1}
\end{figure}

To understand how this comes about, let us first look at the linear (single-plasmon) absorption. Assuming that the graphene plane is perpendicular to the $z$ axis, the plasmon field can be described by the scalar potential $\varphi({\bf r},z,t)=\frac{1}{2}\varphi_pe^{i{\bf q}\cdot{\bf r}-q|z|}e^{-i\omega t} + c.c.$, where ${\bf r}=(x,y)$. The electric field is then given by ${\bf E}=-\mbox{\boldmath $\nabla$}\varphi$, with the amplitude $|E_p|=q|\varphi_p|$, for both in-plane and out-of-plane components.
The interaction of the plasmon field with the electrons can be described by the Hamiltonian $H_p=-\frac{1}{2}e\varphi_p \sum_{j} e^{i{\bf q}\cdot\hat{\bf r}_j}e^{-i\omega t}+c.c.$, where $\hat{\bf r}_j$ is the position operator acting on the $j^{th}$ electron.
To calculate linear absorption we can write the dissipated power as $P=\sum_{n}\hbar\omega \frac{d w_{n0}^{(1)}}{dt}$, where
\begin{equation}
\frac{d w_{n0}^{(1)}}{dt}=\frac{2\pi}{\hbar} |\langle n|H_p|0\rangle|^2
\delta(E_n-E_0-\hbar\omega)
\label{Fermi_golden_rule_linear}
\end{equation}
is the Fermi golden rule for the probability that the system will absorb one plasmon quantum of energy $\hbar\omega$. Here $|n\rangle$ is the many-body excited state of momentum $\hbar{\bf q}$ and energy $\hbar\omega$, with respect to the ground state $|0\rangle$, and we assume that the system is at zero temperature \cite{SM}.
To quantify absorption we can look at the dissipation rate,
\begin{equation}
\gamma=\frac{P}{W}.
\label{dissipation_rate}
\end{equation}
Here $W=\frac{1}{2}\int d^2 r \rho({\bf r})\varphi({\bf r})$ is the total electrostatic energy stored in the system, which is given by $W=\frac{1}{2}\bar{\varepsilon}_r\varepsilon_0|E_p|^2 A/q$, where $A$ is the surface area of the graphene flake. Finally, the relevant figure of merit is the plasmon quality factor $Q=\omega/\gamma$.

From Eq. (\ref{Fermi_golden_rule_linear}) we see that the absorption process consists of a sum over all events where the plasmon can excite a single e-h pair. Conservation of energy and momentum require that $\hbar\omega=E_{n_2 \bf k+q}-E_{n_1 \bf k}$; however, the Pauli principle allows this process only above the threshold condition of $\hbar\omega>2E_F-\hbar v_F q$ (see the gray area in Figure \ref{fig1} (a)).
We can calculate the linear absorption by using the basis of Dirac electron wavefunctions in graphene \cite{SM}. As an example, at energy $\hbar\omega=1.5 \; E_F$, the dissipation rate $\gamma^{(1)}/\omega=1.1$ is so high ($Q=0.9$) that the plasmon is not a well-defined excitation. Below this threshold the Pauli principle blocks the absorption process and the plasmon is a long-lived quasi-particle. However, if we increase the plasmon field, higher-order~(nonlinear) absorption must be accounted for as well.

We note that this simple calculation of the linear absorption gives the same result as from the RPA \cite{SM}. Encouraged by this fact, we proceed to calculate the nonlinear, two-plasmon absorption by writing $P=\sum_{n} 2\hbar\omega\frac{d w^{(2)}_{n0}}{dt}$ and using the Fermi golden rule for the probability that the system absorbs two plasmon quanta:
\begin{equation}
\frac{d w^{(2)}_{n0}}{dt}=\frac{2\pi}{\hbar}
\left| \sum_m \frac{\langle n|H_p|m\rangle \langle m|H_p|0\rangle} {E_m-E_0-\hbar\omega} \right|^2
\delta(E_n-E_0-2\hbar\omega).
\label{Fermi_golden_rule_nonlinear}
\end{equation}
Alternatively one could perform third order expansion of the single particle density matrix including the screening fields consitently in every order of the expansion \cite{Ehrenreich1959}. Such a calculation yields additional terms that contribute to the dissipation, arising from higher-order screening correlations. However, at high fields, when the nonlinear absorption is large, the screening process will be less effective and the simple calculation (\ref{Fermi_golden_rule_nonlinear}) should give a good estimate of the absorption. By evaluating expression (\ref{Fermi_golden_rule_nonlinear}) we get the two-plasmon absorption rate
\begin{equation}
\frac{\gamma^{(2)}}{\omega}=F^{(2)}(\omega)
\left| \frac{E_p}{E_e}\right|^2,
\label{two_plasmon_absorption_rate}
\end{equation}
where $F^{(2)}$ is a dimensionless function of plasmon frequency, which is given in the Supplemental Material \cite{SM} and shown in Figure \ref{fig1} (b). It is straight forward to show that the Pauli principle allows the two-plasmon absorption only above the threshold $\hbar\omega>E_F-\hbar v_F q$ (see the red area in Figure \ref{fig1} (a)). Then if we look at the energy $\hbar\omega=E_F$, single plasmon can not excite an e-h pair but it can decay via two-plasmon absorption. At this energy $F^{(2)}=1.6$ and we see that intrinsic field $E_e$ sets a natural scale for the nonlinear effects. Specifically at the field $E_p=E_e$ the two-plasmon absorption rate $\gamma^{(2)}/\omega=1.6$ is so large that plasmon is not a well defined excitation. In fact at this intrinsic field we expect that the whole perturbation theory should fall apart. 
Indeed at the same energy $\hbar\omega=E_F$ and field strength $E_p=E_e$ we obtain the three-plasmon absorption rate $\gamma^{(3)}/\omega=0.7$ \cite{SM}; which is about $40\%$ of the two-plasmon absorption rate, signalling the breakdown of the perturbation theory. Moreover, at the same field strength but lower energy $\hbar\omega=0.6 \; E_F$ (in the yellow area in Figure \ref{fig1} (a), where two-plasmon absorption is forbidden by the Pauli principle) we obtain the three-plasmon absorption rate $\gamma^{(3)}/\omega=6.5$ again showing that plasmon can not oscillate at the intrisic field $E_e$. 

\begin{figure}
\centerline{
\mbox{\includegraphics[width=0.45\textwidth]{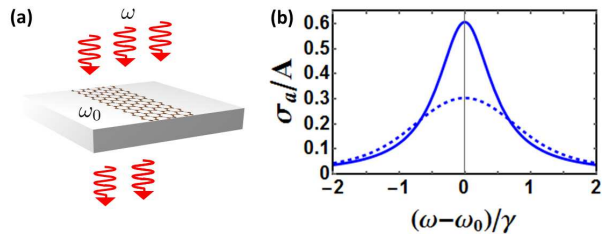}}
}
\caption{
(a) Transmission spectroscopy of a graphene nano-ribbon. (b) Absorption cross section normalized to the surface area of the ribbon. Solid line stands for the low pump $I\ll I_s$, and dashed line for the high pump intensity $I_s=6$ kW/cm$^2$. The saturation of absorption is caused by two-plasmon absorption inside of the ribbon. The ribbon width is $D=25$ nm, doping $n=10^{12}$ cm$^{-2}$, and resonance frequency $\omega_0=2\pi\times$28 THz.
}
\label{fig2}
\end{figure}
These effects could be easily observed in experiments involving extended graphene as the broadening of the plasmon linewidth with increasing field amplitude. However, we will now show that much more exciting behaviour can be seen by using graphene nano-structures to provide a resonance and field enhancement for the plasmon modes. 
The response of graphene nano-structure can be described by the polarizability \cite{Fang2013,Abajo2014}:
\begin{equation}
\alpha(\omega)=D^3\frac{G}
{\frac{K}{\bar\varepsilon_r}-
\frac{4\pi\varepsilon_0 i \omega D}{\sigma(\omega)}}.
\label{polarizability_nanostructure}
\end{equation}
Particularly, in the case of a nano-ribbon, $K=16$ and $G=L/D$, where $D$ is the width and $L\gg D$ is the length of the ribbon~\cite{Abajo2014}. To describe the plasmon resonance we can use the simple Drude model of the surface conductivity $\sigma(\omega)$ given by Eq. (\ref{Drude}). However, now we must include the total damping rate $\gamma_t=\gamma+\gamma_{nl}$, which contains both the linear term ($\gamma$) like impurity or phonon scattering, and the nonlinear term ($\gamma_{nl}$) like two- or three-plasmon absorption. 
Specifically, to produce a resonance at frequency $\omega_0$, we require a ribbon of width: $D=\frac{K}{4\pi^2}\lambda_p\approx\lambda_p/2$, where $\lambda_p=2\pi/q_p$ is the plasmon wavelength in the extended graphene given by Eq. (\ref{plasmon_dispersion}). We are primarily interested in the regime $\hbar\omega_0=E_F$ where a single plasmon can not excite an e-h pair, but two-plasmon absorption is allowed. For a typical doping $n=10^{12}$ cm$^{-2}$~($E_F=0.12$ eV), the corresponding free-space wavelength is $\lambda_0=2\pi c/\omega_0=10.6 \; \mu$m, plasmon wavelength $\lambda_p=62$ nm,  and a ribbon width $D=25$ nm.

The absorption cross-section is given by
$\sigma_{a}(\omega)
=4\pi\frac{\omega}{c} \mbox{Im}\alpha(\omega)
=\frac{3}{4\pi}\lambda_0^2
\frac{\gamma_r \; \gamma_t/2}{(\omega_0-\omega)^2+\gamma_t^2/4}$,
where we have introduced the radiative decay rate of the ribbon $\gamma_r=\frac{16\pi^3}{3}\frac{\bar\varepsilon_r G}{K}\frac{D^3}{\lambda_0^3}\omega_0$~\cite{Abajo2014,Thong2012}. 
To estimate impurity or phonon scattering we can use measurements of direct current mobility $\gamma=\frac{e v_F}{\mu\hbar\sqrt{\pi n}}$ \cite{Jablan2009}. For typical graphene mobilities of $\mu=10^4$ cm$^2$/Vs \cite{Novoselov2004}, we have $\gamma=8.6$ THz, while $\gamma_r=6.1$ GHz, even for a ribbon of length $L=100 \; D$. Since $\gamma_r\ll\gamma$, graphene nano-structures primarily act as absorbers, while the absorbed power can easily be detected via the reduction of light transmitted across the ribbon \cite{Ju2011,Yan2012} (see Figure \ref{fig2} (a)).

Dissipated power is $P=I\sigma_{a}=\gamma_tW$, where  $W$ is the total energy, and $I=\frac{1}{2}\varepsilon_0 c |E|^2$ is the incident light intensity. We can then estimate the plasmon field inside of the ribbon by using the result for the extended graphene
$W=\frac{1}{2}\bar{\varepsilon}_r \varepsilon_0 |E_p|^2 A/q_p$, where $A=LD$ is the ribbon area. We obtain:
$|E_p|^2=\frac{1}{2}\frac{\omega_0^2}{(\omega_0-\omega)^2+\gamma_t^2/4}|E|^2$, where the total damping rate $\gamma_t=\gamma+\gamma_{nl}$ itself depends on the plasmon field through the nonlinear damping term $\gamma_{nl}\propto |E_p|^2$ given by Eq. (\ref{two_plasmon_absorption_rate}). 
Then by increasing the intensity, there is an increase in the total damping rate, and a decrease of the absorption cross section (see Figure \ref{fig2} (b)). In particular, at intensity $I_s=6$ kW/cm$^2$, we obtain $\gamma_{nl}=\gamma$ on resonance, 
which reduces the absorption cross section by a factor of 2. This corresponds to an input power of only 2 mW for a laser focused to a diffraction-limited spot size $(\lambda/2)^2$, which would induce negligible heating of the graphene flake \cite{SM}.
On the other hand, extended graphene can also be used as a saturable absorber, due to Pauli blocking, but the required intensities are 4 orders of magnitude higher, $I_{s}=0.3$ GW/cm$^2$ \cite{Sun2010}. 
Moreover, by lowering the doping we can further reduce the saturation intensity of the ribbon, while the resonance frequency would fall into the terahertz spectrum. 

Especially interesting is the case of a nano-disk where we can localize the entire field to an extremely small volume $\approx \lambda_p^3$. The polarizability of a disk can also be described by the expression (\ref{polarizability_nanostructure}), if we now take $D$ as the disk diameter, $K=12.5$ and $G=0.65$ \cite{Fang2013}. To produce a resonance at energy $\hbar\omega_0=E_F$ and doping $n=10^{12}$ cm$^{-2}$, requires a disk diameter  $D=20$ nm. This yields a radiative decay rate of $\gamma_r=24$ MHz, while the other parameters are the same as in the case of a ribbon (above).

We can now estimate the electric field amplitude associated with a single quantized plasmon by writing $W=\hbar\omega_0$, and using the result for the extended graphene $W=\frac{1}{2}\bar{\varepsilon}_r \varepsilon_0 |E_p|^2 A/q_p$, where $A=\pi D^2/4$ is the disk area. This gives a remarkable result for the field amplitude,
\begin{equation}
\left|E_p^Q \right|= \sqrt{\frac{2^8\pi^4}{K^2}\frac{\hbar\omega_0}{\varepsilon_0\bar\varepsilon_r\lambda_p^3}}
\approx 2\cdot 10^7 \;  {\mbox{V/m}}.
\label{E_field_quantum}
\end{equation}
In other words, the field of a single quantized plasmon is of the same order of magnitude as the intrinsic field $E_e$! Then, for two plasmons the damping rate due to two-plasmon absorption would be so high ($\gamma^{(2)}/\omega_0=5.6$) that the resonance peak would completely disappear, leading to a plasmon blockade effect. 

To quantify this effect we adopt a density matrix approach,
$\frac{d\rho}{dt}=\frac{i}{\hbar}\left[\rho,H\right]+\mathcal{L}\left[\rho\right]$. Here the system Hamiltonian is given by
$H=\hbar\omega_0 N-\frac{\hbar\Omega}{2}(a e^{i\omega t}+a^+e^{-i\omega t})$, where $N=a^+a$ is the plasmon number operator, and $\Omega=\sqrt{\frac{3}{2\pi}\frac{\lambda_0^2 I}{\hbar\omega_0}\gamma_r}$ is the Rabi frequency describing the interaction with the incident field of intensity $I$. The Liouvillian $\mathcal{L}=\mathcal{L}^{(1)}+\mathcal{L}^{(2)}$  consists of two terms that characterize the linear and nonlinear dissipation, respectively, $\mathcal{L}^{(N)}\left[\rho\right]=\frac{\gamma^{(N)}}{2}(2a^N\rho a^{+N}-a^{+N}a^N\rho-\rho a^{+N}a^N)$ \cite{Scully}, where $\gamma^{(1)}=\gamma$ and $\gamma/\omega_0=0.05$, so that $\gamma^{(1)}\ll\gamma^{(2)}$.
Under conditions of weak external driving, the dynamics of the (infinite-dimensional) density matrix can effectively be truncated to a few-excitation manifold~\cite{Scully}. Specifically at $\Omega\ll\gamma^{(2)}$ the steady-state population of the excited state $|2\rangle$ (containing two plasmons) is extremely weakly populated, $\rho_{22}=\frac{2\rho_{11}}{1+(2\gamma^{(2)}/\Omega)^2}$, and disk effectively behaves as a two-level system.

\begin{figure}
\centerline{
\mbox{\includegraphics[width=0.45\textwidth]{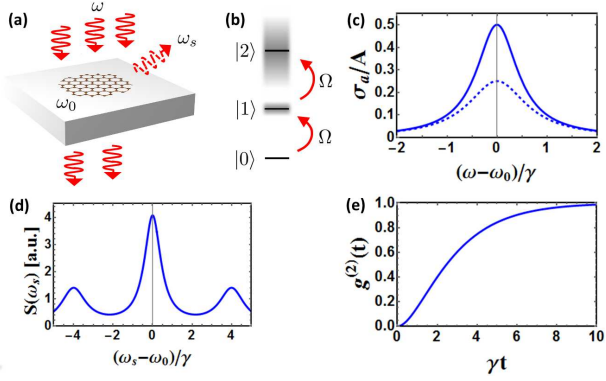}}
}
\caption{
(a) Resonance fluorescence of a graphene nano-disk. (b) Quantized plasmon energy levels in a disk. Two-plasmon absorption induces a huge linewidth of the doubly-excited state $|2\rangle$, which makes it extremely difficult to populate this state and effectively turns the disk into a two-level system.
(c) Absorption cross-section normalized to the disk surface. The solid line stands for the low pump field ($\Omega\ll \gamma$), and the dashed line for the strong pump ($\Omega=\gamma/\sqrt 2$). (d) Scattered spectrum on resonance ($\omega=\omega_0$) in the strong pump regime ($\Omega=4\gamma$),
clearly showing Rabi sidebands at $\omega_0\pm\Omega$. (e) The second-order correlation function $g^{(2)}(t)$ of the scattered field for weak pump intensities, showing an anti-bunching effect $g^{(2)}(0)\ll 1$ characteristic of a two-level system.
}
\label{fig3}
\end{figure}

The absorption cross-section is then given by $\sigma_{a}
=\frac{\hbar\omega_0\Omega}{I}\mbox{Im}\rho_{10}
=\frac{3}{4\pi}\lambda_0^2
\frac{\gamma_r \; \gamma/2}{(\omega_0-\omega)^2+\gamma^2/4+\Omega^2/2}$,
which saturates at intensity $I_s=50$ kW/cm$^2$ ($\Omega=\gamma/\sqrt 2$).
Like in the case of the ribbons one finds that the radiative damping rate is negligible, $\gamma_r\ll \gamma$, and the disk primarily acts as an absorber. However, the weakly scattered light will now show interesting spectral properties, since it is well-known that a two-level system behaves as a strong frequency mixer \cite{Scully}. To substantiate this, let us look at the power spectrum of the scattered light, $S(\omega_s)=\frac{1}{\pi}\mbox{Re}\int_0^\infty \langle a^+(0) a(t) \rangle e^{i\omega_s t}dt$. By using the quantum regression theorem \cite{Scully} we obtain the steady-state two-time correlation function on resonance, $\langle a^+(0) a(t) \rangle=\frac{1}{4}\left( e^{-\gamma t/2}+\frac{1}{2}e^{-3\gamma t/4}e^{-i\Omega t}+\frac{1}{2}e^{-3\gamma t/4}e^{i\Omega t}\right)e^{-i\omega t}$, where we have assumed the strong-pump regime $\Omega\gg \gamma/4$, but also $\Omega\ll\gamma^{(2)}$ so that the disk still behaves as a two-level system.
The scattered spectrum consists of three peaks; one at the laser frequency $\omega_0$ and two at the Rabi sidebands $\omega_0\pm\Omega$. In Figure \ref{fig3} (d) we plot the case of  $\Omega=4\gamma$, while $\Omega/\gamma^{(2)}=0.04$ so that the disk still behaves as a two-level system up to an excellent approximation. A peculiar feature of this result is that the system produces frequency mixing~(characteristic of dispersive nonlinearities) starting only from the two-plasmon absorptive nonlinearity.

The two-level nature of the system is especially reflected in the second-order correlation function $g^{(2)}(t)=\langle a^+(0) a^+(t)a(t) a(0) \rangle/\langle a^+(0)a(0)\rangle^2$, which describes the probability of detecting a second scattered photon at time $t$ given a detection event at $t_0=0$.
In steady-state, low pump regime $\Omega\ll\gamma$, it is straightforward to show that $g^{(2)}(0)=\left( 1+\gamma^{(2)}/\gamma\right)^{-2}=7\cdot 10^{-5}$. Thus, the disk exhibits an almost perfect anti-bunching effect characteristic of an ideal two-level system (to compare, $g^{(2)}(0)=1$ in the absence of nonlinearities). The temporal duration of this anti-bunching dip around zero time delay $t=0$ is given by $\sim 1/\gamma$ (in particular, within a two-level approximation $g^{(2)}(t)=1+e^{-\gamma t}-2e^{-\gamma t/2}$ \cite{Scully}), after which the system returns to a stationary value of $g^{(2)}(t)=1$ as illustrated in Figure \ref{fig3} (e). The sub-Poissonian nature of the scattered light, $g^{(2)}(0)<g^{(2)}(t)$, is a distinctly non-classical feature that reflects the inability of a two-level system to emit two excitations simultaneously \cite{Mandel1986}.

While we have focused primarily on the absorptive nonlinearity, graphene will show also dispersive nonlinearities such as the Kerr effect~\cite{Boyd}. Such a mechanism would in principle result in a shift in the resonance peak, in addition to the saturation of absorption in the graphene nano-ribbon. On the other hand, we do not expect that this would change our results in the case of a nano-disk, since this dispersive nonlinearity also leads to a plasmon blockade effect \cite{Gullans2013}.
Interestingly, the estimated value of the dispersive interaction strength~\cite{Gullans2013} is two orders of magnitude lower than the two-plasmon absorption rate.

In conclusion, we have shown that graphene possesses a strong nonlinear response in the form of multi-plasmon absorption, which leads to the saturation of absorption in graphene nano-ribbons, and the plasmon blockade effect in graphene nano-disks.

The authors would like to acknowledge the support of the European Commission and the Croatian Ministry of Science, Education and Sports Co-Financing Agreement No. 291823, Marie Curie FP7-PEOPLE-2011-COFUND NEWFELPRO project GRANQO, and FP7-ICT-2013-613024-GRASP.

\pagebreak
\widetext

\begin{center}
\textbf{\large Supplemental Material for "Multi-plasmon absorption in graphene"}
\end{center}
\vspace{10mm}

\setcounter{equation}{0}

\renewcommand{\theequation}{S\arabic{equation}}
\renewcommand{\bibnumfmt}[1]{[S#1]}
\renewcommand{\citenumfont}[1]{S#1}

The electron-plasmon interaction can be described by a Hamiltonian:
$H_p=-\frac{1}{2}e\varphi_p(\omega,q)\hat\rho_q^+ e^{-i\omega t}+c.c.$,
where $\varphi_p$ is the plasmon potential at frequency $\omega$ and wavevector $q$, while $\hat\rho_q=\sum\limits_j e^{-i{\bf q}\cdot\hat{\bf r}_j}$ is the Fourier transform of the many-particle density operator
$\hat\rho({\bf r})=\sum\limits_j \delta({\bf r}-\hat{\bf r}_j)$ \cite{SM_Pines}.
To calculate $N$-plasmon absorption we can write the dissipated power as: $P^{(N)}=\sum\limits_{n}N\hbar\omega \frac{d w_{n0}^{(N)}}{dt}$, and use the Fermi golden rule to find the probability of absorbing $N$ plasmon quanta \cite{SM_Sakurai}:
\begin{equation}
\frac{d w_{n0}^{(1)}}{dt}=\frac{2\pi}{\hbar} |\langle n|H_p|0\rangle|^2
\delta(E_n-E_0-\hbar\omega)
\label{SM_Fermi_golden_rule_1plasmon}
\end{equation}

\begin{equation}
\frac{d w^{(2)}_{n0}}{dt}=\frac{2\pi}{\hbar}
\left| \sum_m \frac{\langle n|H_p|m\rangle \langle m|H_p|0\rangle} {E_m-E_0-\hbar\omega} \right|^2
\delta(E_n-E_0-2\hbar\omega)
\label{SM_Fermi_golden_rule_2plasmon}
\end{equation}

\begin{equation}
\frac{d w^{(3)}_{n0}}{dt}=\frac{2\pi}{\hbar}
\left| \sum_{m,l}
\frac{\langle n|H_p|m\rangle \langle m|H_p|l\rangle \langle l|H_p|0\rangle} {(E_m-E_0-2\hbar\omega)(E_l-E_0-\hbar\omega)} \right|^2
\delta(E_n-E_0-3\hbar\omega)
\label{SM_Fermi_golden_rule_3plasmon}
\end{equation}
Here $|n\rangle$ is the many-body excited state of momentum $N\hbar\bf q$ and energy $N\hbar\omega$, with respect to the ground state $|0\rangle$, and we assume that the system is at zero temperature.
In the most simple description of plasmons, we factorize our many-body state into the product of single-particle states which evolve in the mean-field, self-consistent plasmon potential $\varphi_p(\omega,q)=-\frac{e}{2\bar{\varepsilon}_r\varepsilon_0 q}
\langle\hat\rho_q\rangle$ \cite{SM_Pines,SM_Wunsch2006,SM_Hwang2007}.

Low energy single-particle states in graphene are described by Dirac wavefunctions
$\psi_{n\bf k}({\bf r})= \langle {\bf r} | n{\bf k} \rangle=
\frac{1}{\sqrt{2A}}
\left( {\begin{array}{c}
n \\
e^{i \varphi_{\bf k}} \\
\end{array}} \right)
e^{i {\bf k}\cdot{\bf r}}$
with linear dispersion $E_{n \bf k}=n\hbar v_F |{\bf k}|$, where $\bf k$ is the electron wavevector, $n=\pm 1$, and we need to include the two-valley and two-spin degeneracy of the ground state \cite{SM_Wallace1954}.
Finally, the damping rate can be written as $\gamma^{(N)}=P^{(N)}/W$ where $W=\frac{1}{2}\bar{\varepsilon}_r\varepsilon_0 q|\varphi_p(\omega,q)|^2 A$ is the plasmon electrostatic energy, and it is straightforward to show that:
\begin{equation}
\frac{\gamma^{(N)}}{\omega}=F^{(N)}(\omega)
\left| \frac{E_p}{E_e}\right|^{2N-2}
\label{SM_gamma_N}
\end{equation}
where $|E_p|=q|\varphi_p|$ is the plasmon field, $E_e$ is the intrinsic electric field given in the main text, and
$F^{(N)}$ are dimensionless functions of plasmon frequency given by:

\begin{equation}
F^{(1)}(\omega)=\frac{e^2}{ \bar{\varepsilon}_r \varepsilon_0 q}
\frac{4\pi}{A}\sum_{n_1n_2{\bf k}}
|\langle n_2 {\bf k+q}|e^{i{\bf q}\cdot{\bf r}}|n_1 {\bf k}\rangle|^2
\delta(E_{n_2 {\bf k+q}}-E_{n_1 {\bf k}}-\hbar\omega)f_{n_1 {\bf k}}(1-f_{n_2 {\bf k+q}})
\label{SM_F1}
\end{equation}

\begin{align}
F^{(2)}(\omega)=
\left( \frac{\omega}{v_F q} \right)^6
\frac{E_F^3}{k_F^2}
\frac{\pi^2}{A}
\sum_{n_1n_3{\bf k}} &
\left| \sum_{n_2}
\frac{\langle n_3 {\bf k+2q}|e^{i{\bf q}\cdot{\bf r}}|n_2 {\bf k+q}\rangle \langle n_2 {\bf k+q}|e^{i{\bf q}\cdot{\bf r}}|n_1 {\bf k}\rangle}{E_{n_2 {\bf k+q}}-E_{n_1 {\bf k}}-\hbar\omega} \right|^2
\nonumber \\
& \times
\delta(E_{n_3 {\bf k+2q}}-E_{n_1 {\bf k}}-2\hbar\omega)
f_{n_1 {\bf k}}(1-f_{n_3 {\bf k+2q}})
\label{SM_F2}
\end{align}

\begin{align}
F^{(3)}(\omega)=
\left( \frac{\omega}{v_F q} \right)^{10}
\frac{E_F^5}{k_F^2}
\frac{3\pi^2}{2^5A}
\sum_{n_1n_4{\bf k}}  &
\left| \sum_{n_2n_3}
\frac{
\langle n_4 {\bf k+3q}|e^{i{\bf q}\cdot{\bf r}}|n_3 {\bf k+2q}\rangle
\langle n_3 {\bf k+2q}|e^{i{\bf q}\cdot{\bf r}}|n_2 {\bf k+q}\rangle
\langle n_2 {\bf k+q}|e^{i{\bf q}\cdot{\bf r}}|n_1 {\bf k}\rangle
}
{(E_{n_3 {\bf k+2q}}-E_{n_1 {\bf k}}-2\hbar\omega)
(E_{n_2 {\bf k+q}}-E_{n_1 {\bf k}}-\hbar\omega)} \right|^2
\nonumber \\
& \times
\delta(E_{n_4 {\bf k+3q}}-E_{n_1 {\bf k}}-3\hbar\omega)f_{n_1 {\bf k}}(1-f_{n_4 {\bf k+3q}})
\label{SM_F3}
\end{align}
Here $A$ is the surface area of the graphene flake, $k_F$ is the Fermi wavevector, $E_F$ is the Fermi energy, and
$f_{n\bf k}=\Theta(E_F-E_{n\bf k})$ is the Fermi-Dirac distribution at zero temperature.
Also we have explicitly written the expression (\ref{SM_F1}) to be evident that $F^{(1)}(\omega)=2\mbox{Im}\varepsilon(\omega,q)$, where $\varepsilon(\omega,q)$ is the dielectric function calculated in the Random Phase Approximation \cite{SM_Wunsch2006,SM_Hwang2007}, while we have simplified expressions (\ref{SM_F2}) and (\ref{SM_F3}) by using the plasmon dispersion relation $q=\frac{2\pi\varepsilon_0\bar\varepsilon_r\hbar^2}{e^2 E_F}\omega^2$.

Note that even at the room temperature $T=300$ K and doping $n=10^{12}$ cm$^{-2}$, $E_F/kT\approx 4.5$ and our calculations based on the zero-temperature approximation will still be qualitatively valid. To be more specific, at energy $\hbar\omega_0=E_F$ where we were looking at two-plasmon absorption, we will also see a small contribution from the single-plasmon absorption. However, this effect will be suppressed by roughly $e^{-\Delta/kT}\approx e^{-2.3}$ where $\Delta\approx E_F/2$ is the energy gap required to enter the regime of single-plasmon absorption (see Figure 1 (a)).

From a technical stand point, one might also worry about severe heating of the graphene flakes at high laser powers. For example, in the case of a graphene nano-disk, the saturation intensity is $I_s=50$ kW/cm$^2$, with the corresponding absorption cross section on resonance: $\sigma_{a}(\omega_0)\approx 0.25A$ (see Figure 3 (c)), which would induce heating of the disk at $I_s\sigma_{a}/A=13$ kW/cm$^2$. However, note that graphene was reported to heat up at a rate $R=3.3$ K cm$^2$/kW when subjected to a high direct current flow, while sitting on a room temperature SiO$_2$ substrate which acted as a heat sink \cite{SM_Freitag2009}.
Thus, even in the continuous wave regime a laser would increase the temperature of the disk by only $\Delta T\approx 40$ K. Moreover in the case of ribbons, the saturation intensity, and the induced heating is even smaller.

\end{document}